# Interfacial mechanism in the anomalous Hall effect of Co/Bi$_2$O$_3$ bilayers

Edurne Sagasta[1], Juan Borge[2], Luis Esteban[1], Yasutomo Omori[3], Martin Gradhand[4], YoshiChika Otani[3,5], Luis E. Hueso[1,6] and Fèlix Casanova[1,6]

[1]CIC nanoGUNE, Donostia-San Sebastian, Basque Country, Spain.
[2]Departamento de Física de materiales, Universidad del País Vasco, Nano-Bio Spectroscopic group, Donostia-San Sebastian, Basque Country, Spain.
[3]ISSP, University of Tokyo, Kashiwa, Chiba, Japan.
[4] H. H. Wills Physics Laboratory, University of Bristol, Bristol BS8 1TL, United Kingdom
[5]RIKEN-CEMS, Wako, Saitama, Japan.
[6]Basque Foundation for Science, IKERBASQUE, Bilbao, Basque Country, Spain.

**Abstract**

*Oxide interfaces are a source of spin-orbit coupling which can lead to novel spin-to-charge conversion effects. In this work the contribution of the Bi$_2$O$_3$ interface to the anomalous Hall effect of Co is experimentally studied in Co/Bi$_2$O$_3$ bilayers. We evidence a variation of 40% in the AHE of Co when a Bi$_2$O$_3$ capping layer is added to the ferromagnet. This strong variation is attributed to an additional source of asymmetric transport in Co/Bi$_2$O$_3$ bilayers that originates from the Co/Bi$_2$O$_3$ interface and contributes to the skew scattering.*

Spin-orbit coupling (SOC), the interaction between the charge and spin degree of freedom of electrons, is the origin of many novel spin-dependent phenomena which are widely studied in the field of spin-orbitronics [1, 2]. Some particularly relevant for applications are the spin-charge current interconversions: The spin Hall effect (SHE) [3, 4] occurs in the bulk of conductors, where SOC acts as an effective magnetic field that deflects the spin-up and spin-down electrons in opposite direction, and the Edelstein effect [5] at Rashba interfaces [6,7] or surface states of topological insulators [8], where SOC generates a spin texture with spin-momentum locking.

In ferromagnetic (FM) metals, the SHE appears alongside the anomalous Hall effect (AHE) which, due to the unbalanced spin population, generates a transverse charge accumulation when a charge current (which is spin-polarized) is applied to the system [9, 10]. Depending on the origin of the SOC, we distinguish between the intrinsic [11] and extrinsic mechanisms [12, 13]. In the first case, SOC is inherent to the band structure of the material and the intrinsic anomalous Hall conductivity ($\sigma_{AH}^{int}$) is determined by the Berry curvature. $\sigma_{AH}^{int}$ thus depends on the crystallographic phase of the FM: for instance, different values were calculated for hcp-Co and fcc-Co by Roman *et al.*, which are in quite good agreement with experimental results [14]. For a given crystallographic phase, $\sigma_{AH}^{int}$ is generally anisotropic, for instance it is different for bcc Fe(001) and bcc Fe(111) [15, 16]. In a system with less symmetries, more complex antisymmetric responses can be observed as the magnetization is changed [14, 16]. Ab-initio calculations suggest that $\sigma_{AH}^{int}$ may also decrease when entering the dirty limit [17]. In the extrinsic case, the electrons feel an effective SOC induced by the presence of impurities in the lattice [9]. Most conventionally it is distinguished between the skew scattering and side jump and the strength of the mechanisms depends on the type of impurity and the host material [9,18,19].

It has been theoretically predicted that the inversion symmetry breaking at the interface of different materials generates giant SOC that can result in extra spin-charge interconversion effects in the bulk [20-23]. This prediction has been evidenced in the results of ab-initio calculations, which show a large enhancement of the spin-charge interconversion, which is not confined to the interface between the two metals [24,25]. In this framework, it is appealing to unveil whether the inversion symmetry breaking introduced when a FM is interfaced with a non-magnetic (NM) material, either metallic or insulating, can affect the AHE. Interestingly, the AHE has been observed to be modified in the presence of metallic interfaces [26,27].

In this work, we study the AHE in Co/$Bi_2O_3$ bilayers for different Co thicknesses, unraveling the role that the interface between Co and $Bi_2O_3$ plays in the AHE of Co. We consider $Bi_2O_3$ an ideal material since (i) due to its insulating nature, we can discard additional effects such as extra magnetoresistances coming from the NM layer, and (ii) a large Rashba coefficient is expected in our Co/$Bi_2O_3$ system, as the work function of Co is similar to that of Cu [28,29]. A strong variation of the AHE is observed by adding the $Bi_2O_3$ capping layer to the Co. The temperature dependence of the AHE allows us to extract the weight of the intrinsic and extrinsic contributions. We observe that the intrinsic contribution is insensitive to the $Bi_2O_3$ capping layer, demonstrating that no Rashba contribution modifies the intrinsic contribution. Interestingly, it decreases with increasing the residual resistivity of Co, as predicted theoretically when the system enters the dirty regime [17]. In contrast, the $Bi_2O_3$ capping layer acts as a scattering source at the interface, with a contribution to the observed skew scattering that decays with the thickness of Co layer.

Co and Co/$Bi_2O_3$ thin films were deposited *in situ* on top of doped-Si/$SiO_2$ (150 nm) substrates. Co was e-beam evaporated at 0.5 Å/s and ~8×10$^{-7}$ Torr and $Bi_2O_3$ was also e-beam evaporated at 0.1 Å/s and ~2×10$^{-6}$ Torr. 100-µm-wide and 780-µm-long Hall bars were patterned by negative photolithography and subsequent ion-milling etching was performed. The thickness of $Bi_2O_3$ is 20 nm for all the Co($t$)/$Bi_2O_3$ bilayers and the thickness of Co, $t$, varies from 10 to 160 nm. The grazing incidence X-ray diffraction spectrum shows, for all the samples, a broad and low peak at ~44.5º that corresponds to (0002) hcp-Co, indicating that the films consist of small grains of hcp-Co with preferential orientation of the c-axis out of plane [30]. We cannot confirm whether other orientations are also present out of plane, as the corresponding peak might be unresolvable. Longitudinal (inset in Fig. 1a) and transverse (inset in Fig. 1b) magnetotransport measurements were carried out using a "dc reversal" technique [31] in a liquid-He cryostat, applying an external magnetic field $H$ and varying temperature $T$.

The longitudinal resistivity, $\rho_{xx}$, as a function of temperature of the Co($t$) reference layers and Co($t$)/$Bi_2O_3$ bilayers overlap, as expected from $Bi_2O_3$ being an insulator. An example is shown by Fig. 1(a) for 10-nm-thick Co. The transverse resistance, $R_{xy}=V_c/I$, is measured in the Co($t$) reference Hall bars and Co($t$)/$Bi_2O_3$ bilayer Hall bars as a function of the external out-of-plane magnetic field at different temperatures. Figure 1(b) shows the case for a Co thickness of 10 nm at 10 K. At $|H_z| \gtrsim 2$ T, where the magnetization of Co is saturated out of plane, there is a linear dependence of $R_{xy}$ with $H$ in both systems, due to the ordinary Hall effect occurring in Co. Namely, the slopes are the same for Co and Co/$Bi_2O_3$, indicating that the current is flowing through Co in both systems and the density of charge carriers does not change from the reference to the bilayer. At $|H_z| \lesssim 2$ T, we evidence the magnetization rotation. Importantly, the jump of the transverse resistance from positive values [when extrapolated to zero from a linear fitting at high positive magnetic fields, $R_{xy}(H_z=0^+)$] to negative values [when extrapolated to zero from a linear fitting at high negative magnetic fields, $R_{xy}(H_z=0^-)$], which is associated to the AHE, varies from the Co reference sample to the sample with the $Bi_2O_3$ capping. For the case shown in Fig. 1(b), a remarkable ~40% decrease is observed. The large variation in the AHE cannot be attributed to a change in $\rho_{xx}$ of Co, which is very close for the two samples [Fig. 1(a)], and, hence, the effect is arising from the presence of the $Bi_2O_3$ capping. This clearly indicates that, in Co(10)/$Bi_2O_3$, in addition to the regular AHE occurring in the bulk of FM, there is an extra contribution to the AHE.

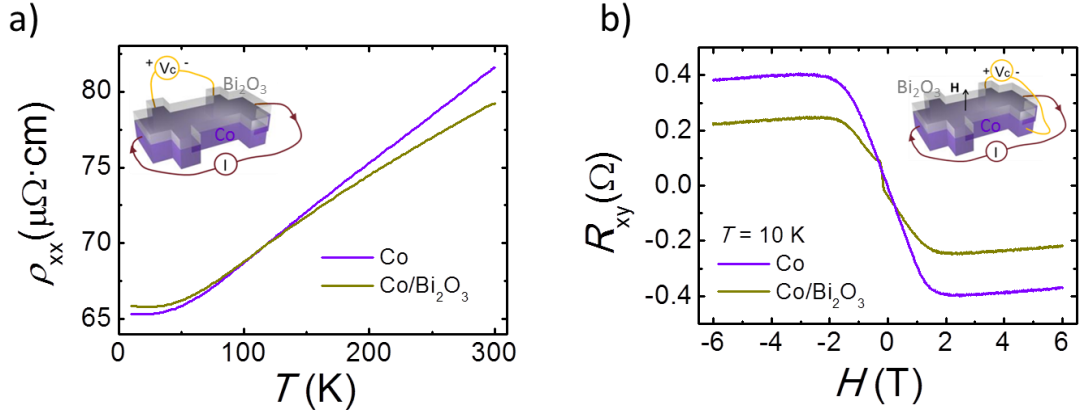

FIG. 1. (a) Temperature dependence of the longitudinal resistivity of Co(10) (purple line) and Co(10)/Bi$_2$O$_3$ (golden line). Inset: Measurement configuration of the longitudinal resistivity. (b) Anomalous Hall effect measurement in Co(10) (purple line) and Co(10)/Bi$_2$O$_3$ (golden line) at 10 K. Inset: Measurement configuration of the transverse resistivity applying out-of-plane magnetic field. The applied current, $I$, is 1 µA in (a) and 10 µA in (b).

We extrapolate $R_{xy}(H=0^+)$ and $R_{xy}(H=0^-)$ values from high magnetic field data and calculate the anomalous Hall resistivity, $\rho_{AH}=t\cdot[R_{xy}(H=0^+)-R_{xy}(H=0^-)]/2$, for both systems at different temperatures. By following the empirical relation for the AHE proposed by Tian et al. [32] that considers both the extrinsic (skew scattering and side jump) and intrinsic contributions to the AHE of Co, we can write the anomalous Hall resistivity as

$$-\rho_{AH} = \sigma_{AH}^{int}\rho_{xx}^2 + \alpha_{AH}^{ss}\rho_{xx0} + \sigma_{AH}^{sj}\rho_{xx0}^2 \qquad (1)$$

where $\sigma_{AH}^{int}$ is the intrinsic anomalous Hall conductivity, $\alpha_{AH}^{ss}$ is the skew scattering angle, $\sigma_{AH}^{sj}$ is the anomalous Hall conductivity that corresponds to side jump contribution and $\rho_{xx0}$ is the residual resistivity. The last two terms represent the extrinsic contribution:

$$-\rho_{AH}^{ext} = \alpha_{AH}^{ss}\rho_{xx0} + \sigma_{AH}^{sj}\rho_{xx0}^2. \qquad (2)$$

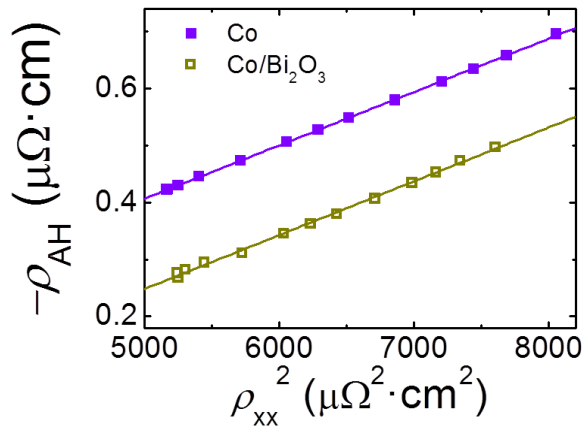

FIG. 2. Anomalous Hall resistivity as a function of the square of the longitudinal resistivity of Co (solid purple squares) and Co/Bi$_2$O$_3$ (open golden squares). Purple solid line (golden solid line) is the fitting of Co (Co/Bi$_2$O$_3$) data to Eq. (1).

Figure 2 shows $\rho_{AH}$ as a function of the square of the longitudinal resistivity of Co for the Co(10) reference sample and the Co(10)/Bi$_2$O$_3$ bilayer. We clearly observe that the slopes of both curves are the same, 93.6±0.6 Ω$^{-1}$cm$^{-1}$ and 94±1 Ω$^{-1}$cm$^{-1}$, respectively, indicating that $\sigma_{AH}^{int}$ is not affected by the $Bi_2O_3$ capping layer on top. However, we obtain a very different extrinsic contribution for each system. $\rho_{AH}^{ext}$ in Co(10)/$Bi_2O_3$ is 3 times larger than in Co(10), suggesting that the Co/$Bi_2O_3$ interface acts as an extra scattering source.

In order to confirm the interfacial origin of the effect, we calculate $\rho_{AH}$ in samples with different Co thicknesses, $t$ = 10, 13, 16, 23, 39, 74, 157 nm, in Co($t$) reference samples and Co($t$)/$Bi_2O_3$ bilayers. The resistivity of Co for the Co($t$) reference samples and Co($t$)/$Bi_2O_3$ bilayers with the same Co thickness is the same, as shown in Fig. 3(a) at 10 K. We observe that the residual resistivity shows a $t^{-1}$ dependence, following the Mayadas and Shatzkes model [33]. Figure 3(b) shows the anomalous Hall resistivity for all the samples with different Co thicknesses, with and without the $Bi_2O_3$ capping layer. Interestingly, the thinnest Co samples show a larger difference between the AHE signals with and without the $Bi_2O_3$ capping, further suggesting that the additional effect has an interfacial origin. We extract the weight of each mechanism ($\sigma_{AH}^{int}$ and $\rho_{AH}^{ext}$) by fitting each individual sample to Eq. (1) as we did previously with $t$ = 10 nm in Fig. 2.

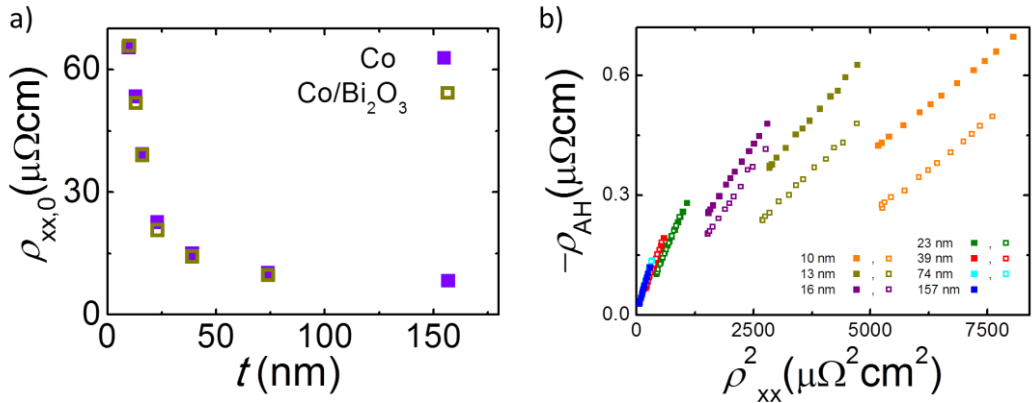

FIG. 3. (a) Residual resistivity of Co as a function of the thickness for the Co reference layers (solid purple squares) and the Co/$Bi_2O_3$ bilayers (open golden squares) at 10 K. (b) Anomalous Hall resistivity as a function of the square of the longitudinal resistivity of Co (solid squares) and Co/$Bi_2O_3$ (open squares) for different Co thicknesses. The applied currents range from 1 to 10 µA in (a) and from 10 to 100 µA in the measurements that gave the results shown in (b).

Figure 4(a) shows the intrinsic anomalous Hall conductivity $\sigma_{AH}^{int}$, obtained from the individual fitting for each sample, as a function of its residual resistivity. There is almost no difference between $\sigma_{AH}^{int}$ obtained for Co($t$)/$Bi_2O_3$ bilayer and Co($t$) reference samples, which is consistent with the result in Fig. 2. Therefore, we confirm that $\sigma_{AH}^{int}$ in Co is independent of the presence of $Bi_2O_3$ capping layer on top. Taking into account that $\sigma_{AH}^{int}$ is a property of the band structure of the material, this result indicates that the $Bi_2O_3$ capping layer is not modifying the band structure of Co.

Interestingly, the same results show that $\sigma_{AH}^{int}$ is modified by the residual resistivity of Co, a feature in principle not expected. For instance, a constant $\sigma_{AH}^{int}$ value of 205 $\Omega^{-1}$cm$^{-1}$ for hcp-Co is reported for a residual resistivity range of 16–42 µΩcm [34], while the $\sigma_{AH}^{int}$ value we obtain for that resistivity range (15–39 µΩcm) decays from 318 to 176 $\Omega^{-1}$cm$^{-1}$. However, our data is in good agreement with the tight-binding calculations performed by Naito *et al*. [17], which show a decay in $\sigma_{AH}^{int}$ as the impurity concentration increases even before entering the dirty limit. They report a value of 341 $\Omega^{-1}$cm$^{-1}$ for Co with a residual resistivity of 5 µΩcm, which decreases to 148 $\Omega^{-1}$cm$^{-1}$ before entering the dirty limit [17]. In our case, we obtain 402±4 $\Omega^{-1}$cm$^{-1}$ for 8.2 µΩcm, which decays to 113.0±0.4 $\Omega^{-1}$cm$^{-1}$ when the residual resistivity increases to 65.3 µΩcm. This agreement suggests that we are experimentally observing the predicted decay of $\sigma_{AH}^{int}$ as the residual resistivity increases in the intermediate (moderately dirty) regime of Co. An alternative explanation could be that the texture of the hcp Co varies with the thickness of Co, going from a *c*−axis orientation of the grains to an *ab*−plane orientation. As reported by Roman *et al.*, $\sigma_{AH}^{int}$ for hcp Co in *c*−axis is 481 $\Omega^{-1}$cm$^{-1}$ and in *ab*−plane is 116 $\Omega^{-1}$cm$^{-1}$ [14], values that would be in agreement with our results. However, we cannot resolve any variation in the texture of our polycrystalline Co films from the x-ray diffraction measurements.

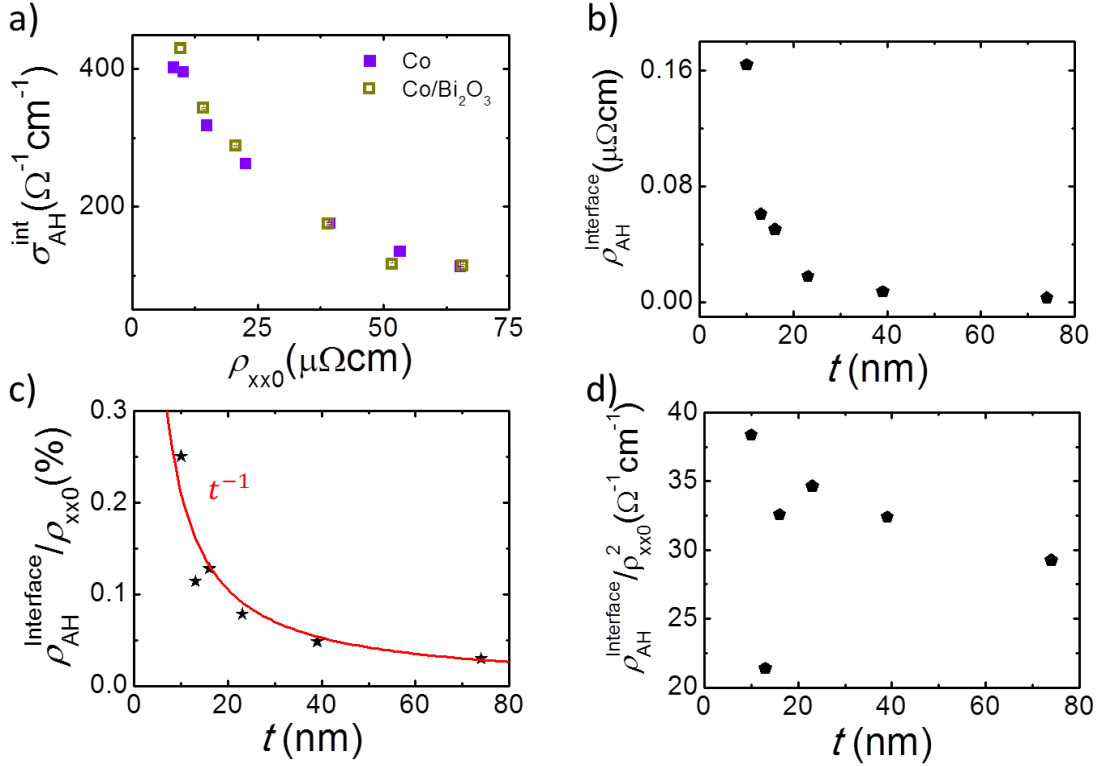

FIG. 4. (a) Residual resistivity dependence of the intrinsic anomalous Hall conductivity of Co for the Co reference layers (solid purple squares) and the Co/Bi$_2$O$_3$ bilayers (open golden squares). (b) Thickness dependence of the additional anomalous Hall resistivity at the interface. (c) Thickness dependence of the ratio of the additional anomalous Hall resistivity at the interface and the residual resistivity of Co. Red solid line is a fit to $t^{-1}$. (d) Thickness dependence of the ratio of the additional anomalous Hall resistivity at the interface and the square of the residual resistivity of Co.

We now turn to the extrinsic contribution $\rho_{AH}^{ext}$, obtained from the individual fitting for each sample. $\rho_{AH}^{ext}$ differs significantly from the reference sample to the bilayer system. We first analyze $\rho_{AH}^{ext}$ in the reference samples, which corresponds to the bulk of Co, in order to disentangle the skew scattering from the side jump contributions. By plotting $-\rho_{AH}^{ext}/\rho_{xx0}$ as a function of $\rho_{xx0}$, we can linearly fit the data to Eq. 2 in order to extract $\sigma_{AH}^{sj}$ from the slope and $\alpha_{AH}^{ss}$ from the intercept. We obtain: $\sigma_{AH}^{sj} = -17 \pm 3 \, \Omega^{-1}\text{cm}^{-1}$ and $\alpha_{AH}^{ss} = 0.04 \pm 0.01$ % for the Co reference samples. This extrinsic contribution from the bulk of the Co layer should also be present in the bilayer system. Therefore, in order to isolate the additional extrinsic contribution that is present only in the bilayer system due to the interface, we subtract $\rho_{AH}^{ext}$ for the corresponding Co reference layer from $\rho_{AH}^{ext}$ of each bilayer, obtaining $\rho_{AH}^{interface}$. $\rho_{AH}^{interface}$ increases when the thickness of the Co layer decreases, as shown in Fig. 4 (b), which points to an interfacial effect. This interfacial extrinsic effect could modify either the skew scattering or the side jump. In order to resolve this question, we plot the characteristic coefficients of each mechanism, $\rho_{AH}^{interface}/\rho_{xx0}$ and $\rho_{AH}^{interface}/\rho_{xx0}^2$ for skew scattering and side jump, respectively, as a function of *t*, see Fig. 4 (c) and (d). Being the effect originated at the interface and the system diffusive, a $t^{-1}$ dependence is expected for the coefficient that is influenced by the interface. Indeed, Fig. 4(c) shows that the ratio between $\rho_{AH}^{interface}$ and $\rho_{xx0}$ follows a $t^{-1}$ dependence, indicating that the interfacial contribution can be written as $\rho_{AH}^{interface} = \alpha_{AH}^{ss,interface} \rho_{xx0}$ where $\alpha_{AH}^{ss,interface}$ shows a $t^{-1}$ dependence. In contrast, the ratio between $\rho_{AH}^{interface}$ and $\rho_{xx0}^2$ does not show any clear dependence with $t$ (see Fig. 4 (d)). Therefore, we conclude that the interface modification, by adding a $Bi_2O_3$ layer on top of Co, results on an interfacial skew scattering contribution of the AHE in Co. Xu et al. reported an interfacial skew scattering in epitaxially grown Ni/Cu metallic bilayers, where $\alpha_{AH}^{ss,interface}$ is constant and does not depend on the thickness of Ni [26]. In contrast to our case, transport in their system is not in the diffusive regime along the thickness because their samples were grown epitaxially and the mean free path is longer than the thickness. A recently reported interface-induced anomalous Hall conductivity [35] is unlikely to be present in our system, given that our samples are polycrystalline.

To conclude, we evidence a variation of up to 40% in the AHE of Co originated by interface modification. The addition of an insulating $Bi_2O_3$ layer on top of Co gives rise to interfacial skew scattering, where the skew scattering angle follows a $t^{-1}$ dependence, characteristic of an interfacial effect. We also observe that the intrinsic spin Hall conductivity of Co is insensitive to the presence of the $Bi_2O_3$ capping layer. $\sigma_{AH}^{int}$ decreases when we increase the residual resistivity in Co, evidencing the influence of the impurities of the bulk of Co on the intrinsic mechanism when the system enters the dirty limit.

## Acknowledgements

This work is supported by the Spanish MINECO under the Maria de Maeztu Units of Excellence Programme (MDM-2016-0618) and under Projects No. MAT2015-65159-R and MAT2017-82071-ERC, and by the Japanese Grant-in-Aid for Scientific Research on Innovative Area, "Nano Spin Conversion Science" (Grant No. 26103002). E.S thanks the Spanish MECD for a Ph.D. fellowship (Grant No. FPU14/03102). Y. Omori acknowledges financial support from JSPS through "Research program for Young Scientists" and "Program for Leading Graduate Schools (MERIT)". M.G. acknowledges financial support from the Leverhulme Trust via an Early Career Research Fellowship (ECF-2013-538).